\documentclass[
reggeiaps,prd,
nofootinbib,
superscriptaddress,
showpacs,ligh
tightenlines,
notitlepage]{revtex4-1}
\usepackage{amsmath}
\usepackage{amssymb}
\usepackage{bm}
\usepackage{color,graphicx}
\usepackage{mciteplus}
\usepackage{textcomp}
\usepackage{slashed}
\usepackage[utf8]{inputenc}

\newcommand{\Sym}{\mathbb{S}}

  \newcommand*\oline[1]{%
      \vbox{%
        \hrule height 0.5pt
        \kern -0.25ex
        \hbox{%
          \kern-0.05em
          \ifmmode#1\else\ensuremath{#1}\fi
          \kern-0.05em
        }
      }
    }

\begin{document}

\title{Quark Orbital Angular Momentum in the MIT Bag Model. }

\author{A.~Courtoy} 
\email{acourtoy@fis.cinvestav.mx}
\affiliation{Departamento de F\'isica, Centro de 
Investigaci\'on y de Estudios Avanzados, Apartado Postal 14-740, 07000 
Ciudad de M\'exico, M\'exico. }
\affiliation{C\'atedra CONACyT}


\author{A.~S.~Miramontes} 
\email{smiramontes@fis.cinvestav.mx}
\affiliation{Departamento de F\'isica, Centro de 
Investigaci\'on y de Estudios Avanzados, Apartado Postal 14-740, 07000 
Ciudad de M\'exico, M\'exico. }



\begin{abstract}
We present the results for the Generalized Transverse Momentum Distribution related to quark Orbital Angular Momentum, {\it i.e.} $F_{14}$, in the MIT bag model. This model has been modified to include the Peierls--Yoccoz projection to restore translational invariance.
Such a modification allows to fulfill more satisfactorily basic sum rules, that would otherwise be less elegantly carried out with the original version. Using the same model, we have calculated the twist-$3$ GPD that corresponds to Orbital Angular Momentum \`a la Ji, through the Penttinen--Polyakov--Shuvaev--Strikman sum rule. Recently, a new relation between the two definitions of the quark Orbital Angular Momentum at the density level has been proposed, which we illustrate here within the model. The sum rule is fulfilled. Still within the framework of the MIT bag model, we analyze the Wandzura--Wilczek expression for the GPD of interest. The genuine quark-gluon contribution is evaluated directly thanks to the equation of motion of the bag, which allows for a direct control of the kinematical contributions to the twist-3 GPD.
\end{abstract}

\maketitle

\section{Introduction}

Since the results of the EMC experiments establishing that the quark spin alone does not satisfy the expected sum rule, there have been numerous studies in the context of hadronic physics to solve this puzzle. 
The Orbital Angular Momentum (OAM) of partons was identified as a key piece of the proton spin puzzle~\cite{Jaffe:1989jz}, a breakthrough that oriented the research focus towards the transverse motion of quarks inside the hadrons. The argument has been eclipsed by the appearance of the total angular momentum sum rules, that would be accessible through Deep exclusive processes~\cite{Ji:1996ek}. However, the resulting definitions for the quark OAM from Jaffe--Manohar~\cite{Jaffe:1989jz} and Ji~\cite{Ji:1996ek} disagreed, what incited to study the discrepancy  in greater details.

In the past few years, the definition of OAM has provoked incessant discussions. 
The outcome of the present dilemma is a classification of the quark and gluon OAM in two main families of spin decomposition, illustrated by the Jaffe--Manohar (JM) and Ji's decomposition respectively, see {\it e.g.}~\cite{Leader:2013jra,Wakamatsu:2014zza} for reviews.

At the hadronic level, the characterization of OAM is embodied through Parton Distribution Functions. While the total angular momentum has been identified as second Mellin moments of Generalized Parton Distributions (GPDs)~\cite{Ji:1996ek}, leading to the beginning of research on a new family of off-forward PDFs, the quark OAM was first indirectly related to GPDs by subtraction, {\it i.e.} $L=J-S$. Penttinen, Polyakov, Shuvaev and Strikman showed that the quark OAM could be directly related to a subleading GPD~\cite{Penttinen:2000dg}. For experimental accessibility reason, that Penttinen--Polyakov--Shuvaev--Strikman (PPSS) sum rule was not pursued as a viable way to OAM until recently~\cite{Courtoy:2013oaa}.

On the other hand, the JM OAM could only recently be ascribed to a distribution function thanks to the emergence of the so-called mother distributions, the Generalized Transverse Momentum Distributions (GTMDs) in turn related to the Wigner distributions~\cite{Meissner:2009ww}. The GTMDs allow to access five dimensions, {\it i.e.} three in the quark momentum and two in impact parameter, supporting the existence of an OAM-like structure $\vec{r}\times\vec{p}$ that disappears when dimensions are integrated out, as demonstrated in Ref.~\cite{Lorce:2011ni}.

The relationship between the two families has been extensively studied, turning into examinations of the two respective distribution functions, {\it i.e.} the subleading GPD and the GTMD. One important result consists in showing that both the JM and Ji OAM could be defined through Wigner average~\cite{Ji:2012sj}, the only difference coming from the choice of the path for the gauge link in the definition of the quark field to include (or not) initial/final state interactions~\cite{Burkardt:2012sd}. 
Another decisive result for the observability of OAM came providing an
explicit link between the two definitions, though with the same choice of gauge link, connecting them through their dependence on partonic intrinsic
transverse momentum. This results in a sum rule at the density level~\cite{Rajan:2016tlg}, a formalism that connects to lattice QCD as shown in the aforementioned reference.
\\

While the theoretical progress have abound towards the understanding of the quark OAM, there exist only few phenomenological examinations, mainly due to the complexity of the DVCS observable identified to contain the required information~\cite{Courtoy:2013oaa}. The same conclusion applies to model calculations. Models offer a framework for the evaluation of new distribution functions and relations. While we are aware of only two model calculation of the subleading GPD $G_2$~\cite{Kiptily:2002nx,Courtoy:2014bea}, the GTMD $F_{14}$ has been evaluated in the light-cone constituent quark model~\cite{Lorce:2011kd}, in a spectator model~\cite{Rajan:2016rlx}, in  a light-front
dressed quark model~\cite{Mukherjee:2014nya}, in a reggeized quark-diquark in a scalar diquark model~\cite{Courtoy:2014bea}, the quark-target model and in an approximated perturbative QCD valid for large quark transverse momenta, in Ref.~\cite{Kanazawa:2014nha}.
\\

In the present paper, we propose an evaluation of both distribution functions of interest in the MIT bag model.
The bag model has shown useful for first calculations of new parton distributions, especially due to its dynamical content. In the original model, free quarks are bound in a spherical cavity. The boundary conditions here play the role of the confining mechanism, somehow mimicking the gluons. While the authors are aware of the shortcomings of calculation within this model, we have the control over the characteristics of the results and the physical content of the latter. In particular, the physical content of the GPD $G_2(x, \xi, t)$ as composed of a Wandzura-Wilczek reducible contribution and a genuine quark--gluon correlation component will be studied here. The analysis is comparable to the treatment of Burkhardt--Cottingham sum rule of Ref.~\cite{Jaffe:1990qh}, in which it is evident that the genuine twist-3 PDFs correspond to quark--boundary correlations.
Also, extensions of the present analysis including a non straight gauge link will be possible within the present model, {\it i.e.} T-odd TMDs have already been calculated in the MIT bag model~\cite{Yuan:2003wk,Courtoy:2008dn,Courtoy:2009pc}.

The calculation is performed in a modified version of the MIT bag model, to include corrections due to the relative motion of the initial and final bag state. Such corrections are especially important for parton distribution defined  from exclusive processes, such as  Generalized Parton Distributions. Though there is no know process for GTMDs so far, it is expected to be exclusive, {\it i.e.} to allow for a momentum transfer $t$.

In Ref.~\cite{Ji:1997gm}, GPDs have been calculated in a boosted bag model including a physically motivated free parameter. As a sanity check, we have compared our results for chiral-even GPDs to the unboosted version proposed in the previous reference. We have also compared the behavior of the electric and magnetic form factors with the same reference as well as with Ref.~\cite{Betz:1983dy}.

An analysis of the OAM, through the Ji sum rule, that is, by subtraction, was performed in the MIT bag model in Ref.~\cite{Scopetta:1999my}.

The paper is organized as follows.  We first recall the formalism for the Generalized Transverse Momentum Distributions, followed in Section III. by a description of the non-perturbative model the calculation will be performed within, namely the MIT Bag model. In Section IV., the results for the quark OAM within the GTMD approach is analyzed using a straight gauge link. In order to compare with the GPD evaluation of the quark OAM, we proceed to the evaluation of the subleading twist GPD, $G_2$, in Section V.. Hence, we show that the result is consistent with the GTMD path to OAM. More interestingly, in Section VI., the newly proposed sum rule relating the density behavior of both distributions of interest is studied, which results to be fulfilled satisfactorily. 

\section{Formalism for GTMDs}

The formalism for Generalized Transverse Momentum Distributions is easily projected for GPDs, {\it i.e.} forward limit, so that we can define a single amplitude and extend it to the GPD case afterwards.
Though it is not ideal in the MIT bag formalism, we will calculate the helicity amplitudes (HA) defined as~\cite{Courtoy:2013oaa},
\begin{eqnarray}
A_{ \Lambda \lambda, \Lambda' \lambda'}= 
&& \int \frac{d z^- d^2\vec{z}_T}{(2 \pi)^3} e^{ixP^+ z^--i\vec{ k}_T\cdot \vec{z}_T} \left. \langle p', \Lambda' \mid {\cal O}_{\lambda' \lambda}(z) \mid p, \Lambda \rangle \right|_{z^+=0}, \nonumber \\
\label{eq:def_gtmd}
\end{eqnarray} 
where in the chiral even sector, 
\begin{eqnarray}
{\cal O}_{\pm \pm}(z) & = & \bar{\psi}\left(-\frac{z}{2}\right) \gamma^+(\mathbb{I} \pm \gamma_5)  \psi\left(\frac{z}{2}\right)\quad.
\end{eqnarray}

The combination of such HAs give the expected structure for the unpolarized and the helicity-related GTMDs,
\begin{eqnarray}
\label{F11}
 F_{11} & = & \left(A_{++,++} + A_{+-,+-} + A_{-+,-+} + A_{--,--} \right)/4\,.\nonumber
\end{eqnarray}

On the other hand, the GTMDs $F_{14}$, and $G_{11}$, that disappears either on the TMD or GPD limit, appears as the following combination
\begin{eqnarray}
\label{F14}
i  \frac{k_1\Delta_2 - k_2 \Delta_1}{M^2} F_{14} 
 & =& \left(A_{++,++} + A_{+-,+-} - A_{-+,-+} - A_{--,--} \right)/4\;,
\label{eq:f14_def}
\end{eqnarray}
$F_{14}$ describes an unpolarized quark in a longitudinally polarized proton, while $G_{11}$ describes a longitudinally polarized quark in an unpolarized proton.

The GTMD-based definition of quark OAM is
\begin{equation}
\label{Lu}
L^{\cal U}_q(x)  = \int d^2 k_T \int d^2 b_T \,  (b_T
\times k_T)_3 {\cal W}^{\, \cal U}(x, k_T, b_T )\quad,
\end{equation}
where ${\cal W}^{\, \cal U}$ is a Wigner distribution corresponding to $F_{14}$ and ${\cal U}$ in
 denotes the gauge link, {\it i.e.}, the Wilson
path-ordered exponential connecting the coordinates $-z/2$ and $z/2$. We work, throughout this paper, with straight gauge link (and appropriate gauge), as opposed to staples/light-cone gauge link, corresponding to what is known as Ji's decomposition of 
angular momentum \cite{Ji:2012sj,Lorce:2012ce,Burkardt:2012sd}.

\section{The MIT Bag Model}

For the evaluation in a quark model, we define the proton states within the Peierls-Yoccoz projection, using the formalism described in Refs.~\cite{Benesh:1987ie, Chang:2012nw}, which takes into account the relative motion of the initial and final bags, restoring the lost translational invariance, shortcoming of the static-bag wave function. Besides the problem related to translational invariance, in principle, one should boost the wave function of the moving nucleon as the usual static-bag wave function is identified with the zero-momentum eigenstate.  This is particularily important when going to the Breit frame, as it is customary due to its convenience for the bag calculations. Here we apply the simple prescription of a boosted wave function and then removing the boost effect, which should account for the change of reference frame~\cite{Ji:1997gm}. The effective momentum transfer through the active quark then reads $\vec{\tilde{\Delta}}=(1-\epsilon_0/M)\, \vec{\Delta}=0.75\, \vec{\Delta}$.

The conventions of kinematics for the calculation in the Breit frame are then
\begin{eqnarray}
&&p^{\mu}=(p^0,\vec{p})=\left( \bar{M}, -\frac{\vec{\Delta}}{2}\right) \quad, \nonumber\\
&&p'^{\mu}=(p'^{0},\vec{p}')=\left(  \bar{M},\frac{\vec{\Delta}}{2} \right)  \quad,\nonumber\\
&&z=(z^0, \vec{z})\quad,\quad  \quad t= -\vec{\Delta}^2\quad,
\nonumber\\
&&\xi=\left(\frac{p-p'}{p+p'} \right)^+ =-\frac{\Delta_z}{2 \bar M}\quad,
\label{eq:quark_sym_kin}
\end{eqnarray}
with $z^+=0$ and ${\bar M}^2=M^2-t/4$. Therefore, the maximum physical value for $\xi$ is $\sqrt{- t}/{2\bar M}$.\\

We can now express the proton state, with the conventions in Eqs.~(\ref{eq:quark_sym_kin}), 
\begin{eqnarray}
\left | p,\Lambda\right\rangle 
&=& \left | p^0, -\frac{\vec{\Delta}}{2}\right\rangle_{\Lambda}  \nonumber\\
&=& \frac{1}{\Phi_3(-\frac{\vec{\tilde{\Delta}}}{2})} \int d^3 a \, e^{-i\frac{\vec{\tilde{\Delta}}}{2}\cdot \vec{a}} \,b_1^{\dagger}(\vec{a})b_2^{\dagger}(\vec{a})b_3^{\dagger}(\vec{a}) 
 \left | EB, \vec{r}=\vec{a}\right\rangle_{\Lambda}  \,,
 %
\end{eqnarray}
where $EB$ stands for empty bag, $p$ for the $4$-momentum of the  proton and $\Lambda$ for its helicity. The annihilation and creation operators are $b_{1'} (\vec{b})$ and $b_1^{\dagger}(\vec{a})$, respectively, of a quark with quantum numbers set $1' (1)$, in a bag centered at $\vec{r}=\vec{b} (\vec{a})$. 
The normalization of the state in the Peierls-Yoccoz projection is fixed by $\Phi_3$, defined as
\begin{eqnarray}
\left |\Phi_n(\vec{p}) \right |^2 &=& \int d\vec{a} \; e^{-i\vec{p}\,\cdot\,\vec{a}} \left [ \int d\vec{x} \varphi^{\dagger} (\vec{x}-\vec{a})\, \varphi(\vec{x})\right]^n\quad,
\label{eqn:defphi}
\end{eqnarray}

The quark fields, $\psi$,  in terms of bag fields, $\varphi$, are defined as,
\begin{eqnarray}
 \psi\left(\frac{z}{2}\right)&=&\sum_{n, \kappa} b_{n, \kappa} (\vec{a})\; \varphi_{n}\left(\frac{\vec{z}}{2}-\vec{a}\right) \, e^{-i\frac{\omega_{n \kappa} z_0/2 }{R_0}}\quad.
\end{eqnarray}
The bag wave function reads
\begin{eqnarray}
\varphi_m(\vec{k})&=&i\, \sqrt{4\pi}\, N\, R_0^3 
 \begin{pmatrix}
 t_0(|\vec{k}|) \chi_m\\ 
   {\vec{\sigma} \cdot \hat{k}}\,t_1(|\vec{k}|)\, \chi_m
  \end{pmatrix} \quad,
  \label{bagwf}
\end{eqnarray}
with the normalization factor $N$
\begin{eqnarray}
N&=&\left( \frac{\omega^3}{2R_0^3\, (\omega-1)\sin^2\omega}\right)^{1/2}\quad;\nonumber
\end{eqnarray}
where $\omega=2.04$ for the lowest mode and $R_0$ is the bag radius. The two last quantities are related through the relation $R_0M_P=4\omega$.
The functions $t_i(k)$ are defined as
\begin{eqnarray}
 t_i(k) &=&\int_0^1 u^2\, du\, j_i	(ukR_0)j_i(u\omega)\quad.
\end{eqnarray}

We introduce an  overall factor due the bag normalization
\begin{eqnarray}
C_{\mbox{\footnotesize bag}}&\equiv&  \frac{ 4\pi \, N^2\, R_0^6}{(2\pi)^3}=\frac{16\omega^4}{ \pi^2\, j_0^2(\omega)(\omega-1)M_P^3}\quad,
\end{eqnarray}
where the $(2\pi)^3$ factor is introduced for further convenience.
\\

Within our framework, the contributions for each separate flavor and color combinations, $(n, n') $ and $(\kappa, \kappa')$, in Eq.~(\ref{eq:def_gtmd}) now reads,
%
%
\begin{eqnarray}
A_{\Lambda \kappa, \Lambda' \kappa'}^{n, n'}&=&(2 \pi)^3\,C_{\mbox{\footnotesize bag}}\,\int \frac{d z^- \, d^2 \vec{z}_T}{(2 \pi)^3} e^{ixP^+ z^--i \vec{ k}_T\cdot \vec{z}_T} \frac{1}{\left|\Phi_3(\frac{\vec{\tilde{\Delta}}}{2})\right|^2}
\int d^3a\;d^3b\; C_{\Lambda\Lambda'}^{n n', \kappa  \kappa'}(\vec{a},\vec{b}) \nonumber\\
&& e^{-i \frac{\vec{\tilde{\Delta}}}{2}\, \cdot\,( \vec{a}+ \vec{b})} \;e^{-i\omega_{n \kappa} z_0/R_0}\, \overline{\varphi}_{n'}\left(-\frac{\vec{z}}{2}-\vec{b}\right) \Gamma  \varphi_{n} \left(\frac{\vec{z}}{2}-\vec{a}\right) \quad,
\label{eq:1step_sym}
\end{eqnarray}
%
%
where $\Gamma$ stands either for $ \gamma^+(\mathbb{I} \pm \gamma_5) $. The flavor--spin coefficients expressed through the $SU(6)$ proton Wave Function are given by
%
%
\begin{eqnarray}
C_{\Lambda\Lambda'}^{n n', \kappa  \kappa'}(\vec{a},\vec{b})&=& \frac{1}{18} \displaystyle\sum_{M\neq N \neq P, Q\neq R \neq S} (-1) \epsilon_{MNP}\epsilon_{QRS}\nonumber\\
&& 
\left\langle 0 \left |
\left(b_{u, \Lambda'}^{b}(M)\, b_{u, \Lambda'}^{b}(N) \, b_{d, - \Lambda'}^{b}(P) \; -\; b_{u,  \Lambda'}^{b}(M)\, b_{u, - \Lambda'}^{b}(N) \, b_{d,  \Lambda'}^{b}(P) \right)\right.\right.\nonumber\\
&& \left. \left.b_{n, \kappa}^{\dagger} (\vec{b}) b_{n, \kappa} (\vec{a}) 
\left( b_{u,\Lambda}^{\dagger\, a}(Q)\, b_{u, \Lambda}^{\dagger\, a}(R) \, b_{d, -\Lambda}^{\dagger\, a}(S) \; -\; b_{u, \Lambda}^{\dagger\, a}(Q)\, b_{u, -\Lambda}^{\dagger\, a}(R) \, b_{d, \Lambda}^{\dagger\, a}(S)  \right)
\right | 0 \right\rangle\quad.
\label{eq:col_spin_coef}
\end{eqnarray}
%
%

The antisymmetrizer ensures the overall proton wavefunction is antisymmetric. No flavor change can occur in the present calculation, so we safely set $n=n'$.
Due to the Dirac structure we will explore in the present paper, only the terms proportional to $b^{\dagger}(\vec{a})b(\vec{b})$, which accounts for the displacement of the bags,  will survive. We use the normalization convention
\begin{eqnarray}
\left \{  b_n^{\dagger}(\vec{a}), b_m(\vec{b})\right \} 
&&= \delta_{nm}\, \int d^3 x \, \varphi^{\dagger}_n (\vec{x}-\vec{a})\, \varphi_m (\vec{x}-\vec{b})\equiv I(\vec{a},\vec{b})\quad.
\label{eq:iab}
\end{eqnarray}

The displacement of the bags is a relative motion, opposed to average center-of-mass motion. So, the distribution will depend on that relative motion of the two bags {\it w.r.t.} one another, {\it i.e.} $\vec{b}-\vec{a}$. 
Using  Eq.~(\ref{eqn:defphi}) and Eq.~(\ref{eq:iab}),
we get
\begin{eqnarray}
 I^2(\vec{a},\vec{b})&=&\int \frac{d\vec{k}_1}{(2\pi)^3}\, e^{i \vec{k}_1\, \cdot\, (\vec{a}-\vec{b})}\, \left |\Phi_2(\vec{k}_1) \right |^2\quad.
\end{eqnarray}

It straightforwardly comes out that the flavor--spin coefficients  factorize into a term proportional to the bag displacements represented by $I(\vec{a},\vec{b})$, and a pure flavor--spin coefficient. Those coefficients are given in Tab.~\ref{tab:col_spin_num}. They correspond the  expected $SU(6)$ coefficients.

%
%
\begin{center}
\begin{table}
\begin{tabular}{ | c| c| }
\hline
$C_{++}^{d, \downarrow\downarrow}=\cfrac{2}{3}$ & $C_{++}^{d, \uparrow\uparrow}=\cfrac{1}{3}$\\[2ex]
$C_{++}^{u, \downarrow\downarrow}=\cfrac{1}{3}$ & $C_{++}^{u, \uparrow\uparrow}=\cfrac{5}{3}$
\\[2ex]
\hline
$C_{--}^{d, \downarrow\downarrow}=\cfrac{1}{3}$ & $C_{--}^{d, \uparrow\uparrow}=\cfrac{2}{3}$\\[2ex]
$C_{--}^{u, \downarrow\downarrow}=\cfrac{5}{3}$ & $C_{--}^{u, \uparrow\uparrow}=\cfrac{1}{3}$\\[2ex]
\hline
$C_{+-}^{u, \kappa'=\uparrow\, \kappa=\downarrow}=\cfrac{4}{3}$ & $C_{-+}^{u, \kappa'=\downarrow\, \kappa=\uparrow}=\cfrac{4}{3}$\\[2ex]
$C_{+-}^{d, \kappa'=\uparrow\, \kappa=\downarrow}=-\cfrac{1}{3}$ & $C_{-+}^{d, \kappa'=\downarrow\, \kappa=\uparrow}=-\cfrac{1}{3}$\\[2ex]
\hline
\end{tabular}
\caption{Flavor--spin coefficients calculated from Eq.~(\ref{eq:col_spin_coef}).}
\label{tab:col_spin_num}
\end{table}
\end{center}

The final expression for the HAs is obtained after integration over $(z^-, \vec{z}_T)$ and $k$'s, carrying out straightforward delta-functions. We end up with the expression%
%
\begin{eqnarray}
A_{\Lambda \kappa, \Lambda' \kappa'}^n
&=&\sqrt{2}\,C_{\mbox{\footnotesize bag}}\frac{ C_{\Lambda\Lambda'}^{n, \kappa \kappa'} }{\left|\Phi_3(\frac{\vec{\tilde{\Delta}}}{2}) \right|^2}\, \,
\left |\Phi_2\left (x{\bar M}-\frac{\omega}{R}, \, \vec{k}_T \right) \right |^2 \nonumber\\
&&\times
\;  \varphi^{\dagger}_{n} \left(x{\bar M}-\frac{\omega}{R}+\frac{\tilde{\Delta}^z}{2}, \, \vec{k}_T+\frac{\vec{\tilde{\Delta}}_T}{2} \right)\; \gamma^0\Gamma\; \varphi_{n} \left(x{\bar M}-\frac{\omega}{R}-\frac{\tilde{\Delta}^z}{2}, \, \vec{k}_T-\frac{\vec{\tilde{\Delta}}_T}{2} \right)
\quad.
\label{eq:final}
\end{eqnarray}
%
From now on, we will use the notation,
%
%
\begin{eqnarray}
\vec{k}'
&=&\left ({k}^z_b-\eta\xi {\bar M}, \vec{ k}_{T}+\frac{\vec{\tilde{\Delta}}_T}{2}\right)\quad ;
\nonumber\\
\vec{k}_3
&=&\left ({k}^z_b+\eta\xi {\bar M}, \vec{k}_{T}-\frac{\vec{\tilde{\Delta}}_T}{2}\right)\quad ;\nonumber
\end{eqnarray}%
%
%
with ${k}^z_b$ being the bag energy condition ${ k}^z_b=x {\bar M}-\omega/R$.

The functions $\Phi_n$ used in the PY projection, are~\cite{Chang:2012nw},
\begin{eqnarray} 
 |\Phi_n({\bf p})|^2 &= \cfrac{2^{4-n}\pi R^3\omega^{n-2}}{|{\bf p}|R(\omega^2 - \sin^2 \omega)^n}\displaystyle\int_0^\omega\! \cfrac{d v}{v^{n-1}}\,  \sin \frac{2|{\bf p}|R v}{\omega}\, T^n(v)\;,\nonumber\\
\end{eqnarray}
with
\begin{eqnarray}
T(v) &=& \Big(\omega \!-\! \cfrac{1-\cos 2\omega}{2\omega} \!-\! v \Big) \sin 2v
 - \Big(\cfrac{1}{2} \!+\! \cfrac{\sin 2\omega}{2\omega}\Big) \cos 2v 
 + \cfrac{1}{2} + \cfrac{\sin 2\omega}{2\omega} - \cfrac{1- \cos 2\omega}{2\omega^2} v^2 
\,.\end{eqnarray}

\section{OAM through GTMD}

In Ref.~\cite{Lorce:2011ni}, it was demonstrated that the quark OAM is related, through Wigner average, to the GTMD $F_{14}$. Though this distribution function was first  ascribed to the Jaffe-Manohar definition of the OAM, as shown in Ref.~\cite{Ji:2012sj}, both definitions can be averaged through Wigner-like distributions and only differ in the choice of the gauge-link. Here, we focus particularly on the straight gauge-link, namely: Ji's definition of OAM.

Using the framework of the MIT bag model described in the previous Section together with the combination of helicity amplitudes Eq.~(\ref{eq:f14_def}), we obtain the following expression for the GTMD under consideration,
\begin{eqnarray}
 i  \frac{{ k}_x{\Delta}_y - { k}_y {\Delta}_x}{M^2}F_{14}^u 
&=&\left( A_{++,++}^u+A_{+-,+-}^u-A_{-+,-+}^u-A_{--,--}^u\right)/4\nonumber\\
&=& i\frac{4}{3}\,C_{\mbox{\footnotesize bag}}\frac{\left |\Phi_2\left (x{\bar M}-\frac{\omega}{R}, \, \vec{k}_T \right) \right |^2  }{\left|\Phi_3(\frac{\vec{\tilde{\Delta}}}{2}) \right|^2}
\frac{\left(\vec{k}'\times \vec{k}_{3}\right)_z}{k_3 k'} t_1(k_3)t_1(k')\,.\nonumber\\
\end{eqnarray}
The particular combination of helicity amplitudes comes from the fact that $F_{14}$ represents unpolarized quarks inside a longitudinally polarized proton. We then straightforwardly get
\begin{eqnarray}
 F_{14}^u(x,\xi, k_T^2, \vec{k}_T\cdot \vec{\Delta}_T, t)
&=&-\frac{4}{3}\,C_{\mbox{\footnotesize bag}}\frac{\left |\Phi_2\left (x{\bar M}-\frac{\omega}{R}, \, \vec{k}_T \right) \right |^2  }{\left|\Phi_3(\frac{\vec{\tilde{\Delta}}}{2}) \right|^2}\, \,\frac{M^2}{k_3 k'} t_1(k_3)t_1(k')\quad,\nonumber\\
\\
&&F_{14}^d=-\frac{F_{14}^u}{4}\quad.
\label{eq:f14_bag}
\end{eqnarray}

Notice that the proportionality of the up and down distribution comes from $SU(6)$ symmetry and the Lorentz structure of the GTMD we are considering, with the flavor-spin coefficient given in Table~\ref{tab:col_spin_num}. As a counter example, the GTMDs $F_{11}$ and $F'_{13}$ mix in the Breit frame, what was already shown in Ref.~\cite{Ji:1997gm} and here below. The results for more GTMDs, including the ones here mentioned, are available in Ref.~\cite{Angel:2016}.
\\

\begin{figure}
\includegraphics[width=8.cm]{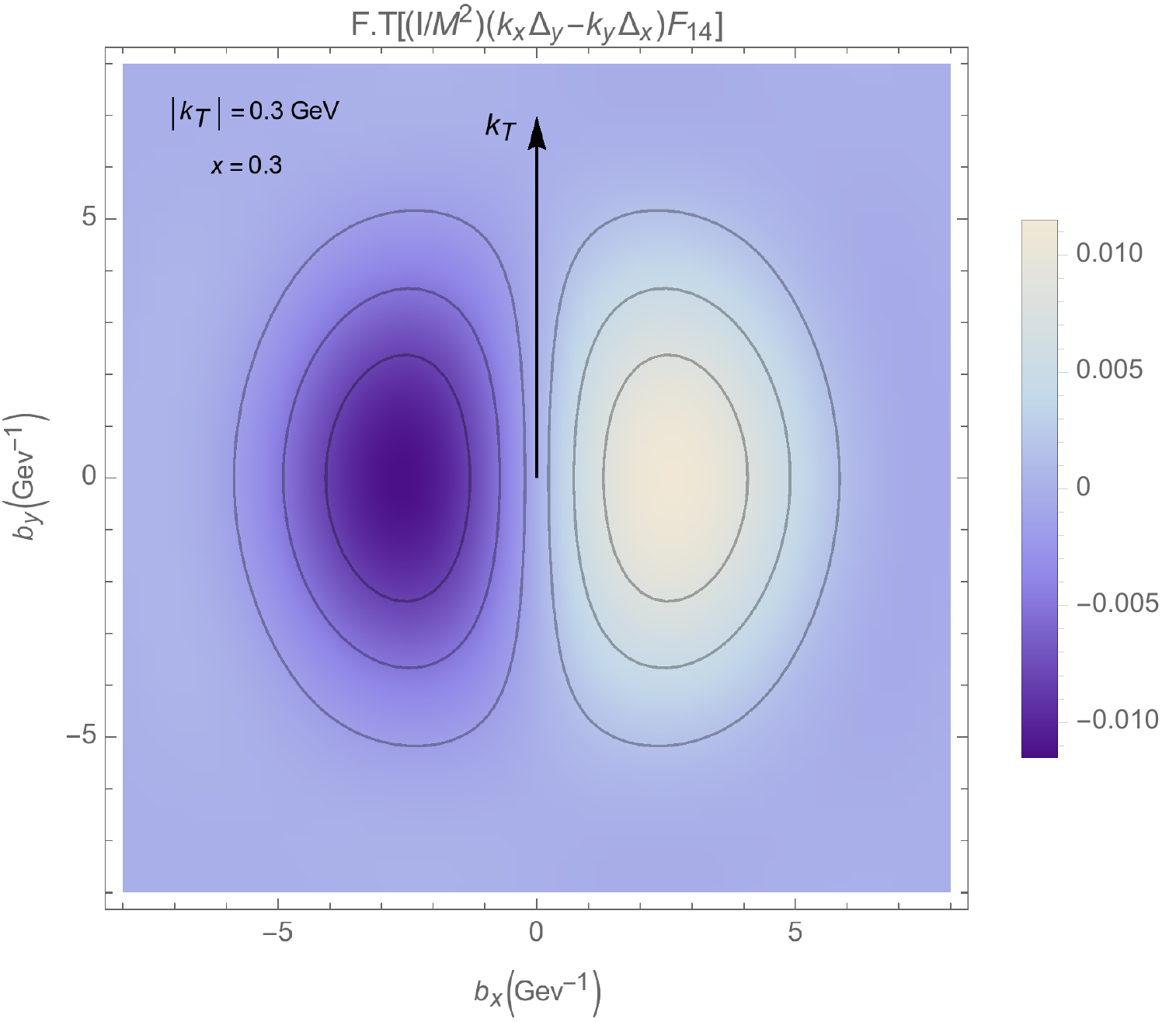}
\includegraphics[width=8.cm]{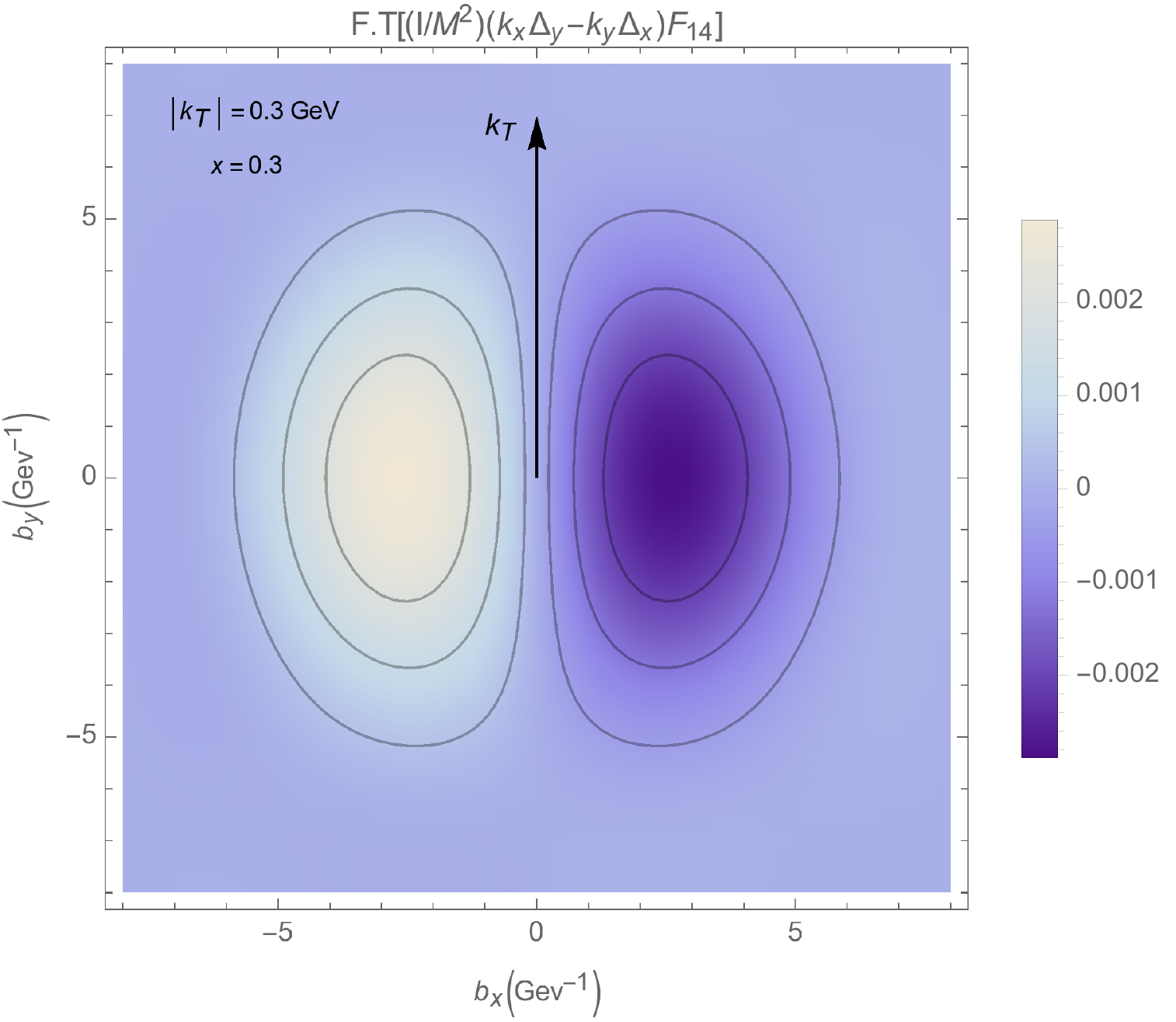}
\caption{Fourier Transform of $F_{14}$ with the PY projection $x=0.3$ and $\vec{k}_T=(0.3$ GeV)$ {\hat e}_y$. Up distribution on the {\it l.h.s} and down on the {\it r.h.s.}. Both are proportional as explained in the text.}
\label{fig:f14_FT}
\end{figure}

In Fig.~\ref{fig:f14_FT}, we show the Fourier transform for the OAM related structure,
\begin{eqnarray}
\mbox{F.T.}\left[ i \frac{k_x\Delta_y-k_y\Delta_x}{M^2} \, F_{14}(x, 0, \vec{k}_T^2, \vec{k}_T\cdot \vec{\Delta}_T,  \vec{\Delta}_T^2)\right]
&=& -\frac{\epsilon^{ij}k_T^i}{M^2}\frac{\partial}{\partial b_T^j} {\cal F}_{14}(x, 0, \vec{k}_T^2, \vec{k}_T\cdot \vec{b}_T,  \vec{b}_T^2)\quad,
\end{eqnarray}
 for fixed values of $|k_T|$ and $x$. The Fourier variable of the momentum transfer corresponds to the impact parameter,~$b_T$. A dipole structure is recognized. The obtained results are in agreement with previous evaluations, {\it e.g.} Ref.~\cite{Lorce:2011kd,Courtoy:2014bea}.
\\

\begin{figure}[h]
\includegraphics[width=8.cm]{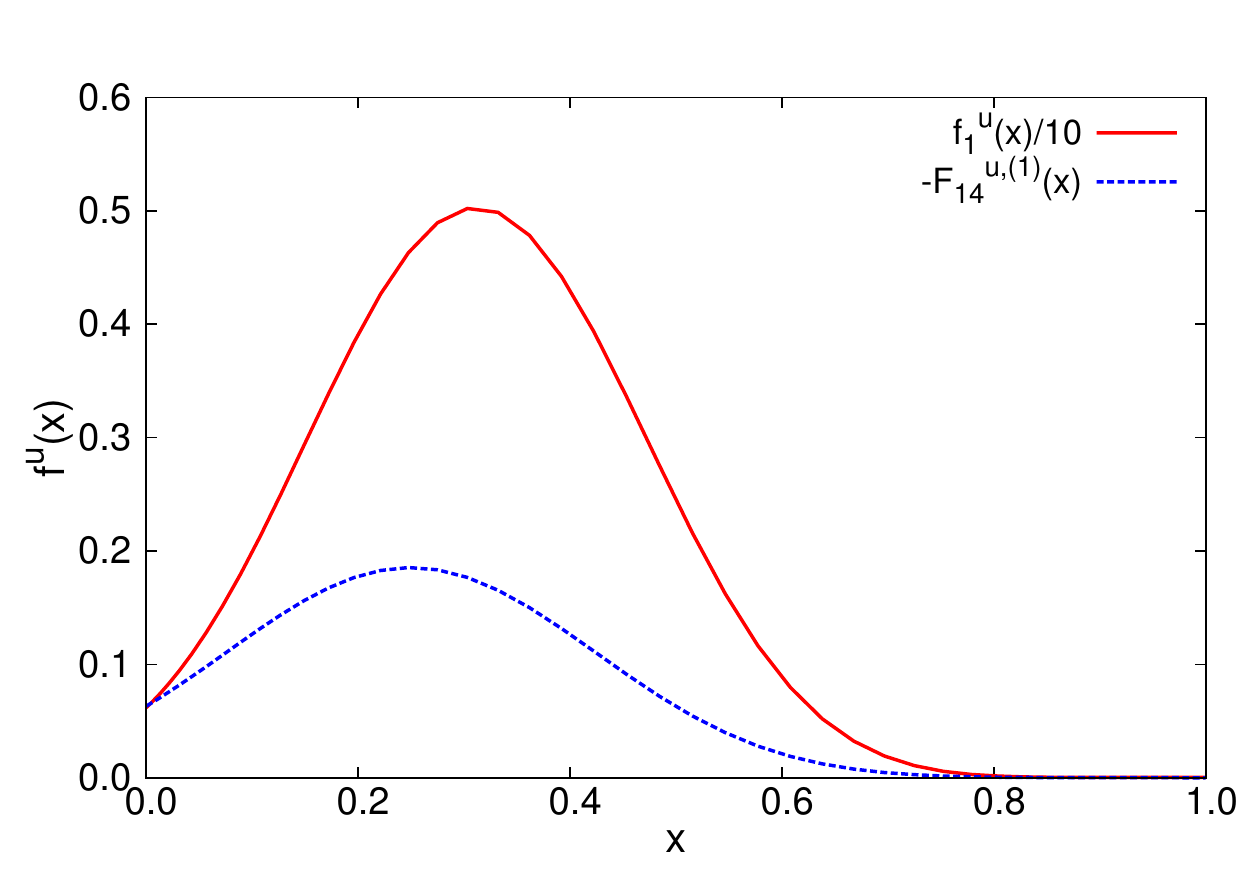}
\caption{The up quark $k_T$-moment of $ F_{14}$, Eq.~(\ref{eq:f14dens}), compared to $f_1^u(x)$ (reduced of a factor 10), both curves with the PY projection.}
\label{fig:f14}
\end{figure}

It is useful to define the $k_T$-moment of $ F_{14}$,
\begin{eqnarray}
F_{14}^{q, (1)}(x, \xi, t)
&=& 2 \int d k_T k_T \int_0^{2\pi} d\phi \sin\phi^2 \frac{k_T^2}{M^2} F_{14}^q(x,\xi, k_T^2, \vec{k}_T\cdot \vec{\Delta}_T, t)\,,\nonumber\\
\end{eqnarray}
where $\phi$ is the angle between $\vec{k}_T$ and $\vec{\Delta}_T$ and reduces to the polar angle of $\vec{k}_T$ in the forward case. In that particular case, we can define the $x$-density,
\begin{eqnarray}
F_{14}^{q, (1)}(x)&=& \int d^2 \vec{k}_T \frac{k_T^2}{M^2} F_{14}^q(x,0, k_T^2, 0,0)\quad.
\label{eq:f14dens}
\end{eqnarray}
We illustrate the result in Fig.~\ref{fig:f14}.

\section{OAM through twist-3 GPD}

Long before the introduction of GTMDs, GPDs were historically proposed to access quark and gluon Angular Momentum (AM)~\cite{Ji:1996ek}. Knowing the spin contribution and the total AM, the OAM can easily be inferred by subtraction. Yet a subleading GPD was shown to be related to the OAM structure \`a la Ji~\cite{Penttinen:2000dg}.
We therefore  turn now to the PPSS sum rule and calculate the twist-$3$ GPD $G_2$~\footnote{It is called $G_3$ in Ref.~\cite{Penttinen:2000dg} and then $G_2$ in Ref.~\cite{Kiptily:2002nx}.},
\begin{eqnarray}
\lim_{t\to 0}\, \int_0^1 dx \, x G_2^q(x,\xi, t)&=&-L^q\quad.
\end{eqnarray}
\\
This sum rule was obtained by subtraction of the structures related to the OAM and the total AM,
\begin{eqnarray}
\lim_{t\to 0}\, \int_0^1 dx \, x G_2^q(x,\xi, t)
&=&\frac{1}{2}\left[\int_0^1 dx g^q_1(x)- \lim_{t\to 0} \int_0^1 dx  x \left(H^q(x,\xi, t)+E^q(x,\xi, t)\right)\right],\nonumber\\
\end{eqnarray}
where the {\it l.h.s.} gives the required quark OAM.
\\

This twist-$3$ GPD can be defined as the $k_T$-integral of twist-$3$ GTMDs~\cite{Meissner:2009ww}. Notice that such integrals are well defined in our model calculation. Switching to the notation of that reference~\cite{Courtoy:2013oaa}, we obtain
\begin{eqnarray}
 -\int dx\, x  G_2(x,\xi, t)
 & =&\int dx\, x \left[\widetilde{E}_{2T}(x,\xi, t) +\left(H (x,\xi, t)+E(x,\xi, t) \right)\right]\,.\nonumber\\
\end{eqnarray}
Namely, the evaluation of the PPSS sum rule requires knowledge on two distribution functions, a twist-3 GPD $\widetilde{E}_{2T}$ and the combination of the twist-2 structure $H+E$.
\\

Let us start with the twist-2 GPD part.
 Working out the nucleon spinor decomposition, we find that, in the Breit frame~\cite{Ji:1997gm},
%
%
\begin{eqnarray}
&&\int \frac{d z^-}{(2 \pi)} e^{ixP^+ z^-}  \langle p', \Lambda' \mid \bar{\psi}\left(-\frac{z}{2}\right) \gamma^+  \psi\left(\frac{z}{2}\right) \mid p, \Lambda \rangle\Bigr|_{\substack{z^+=0\\\vec{z}_{\perp}=0}}\nonumber\\
&&=H(x,\xi, t) \bar{u} (P', \Lambda') \gamma^+u(P, \Lambda)+E(x,\xi, t) \bar{u} (P', \Lambda') \frac{i \sigma^{+\Delta}}{2M}u(P, \Lambda)
\quad, \nonumber \\
\nonumber \\
&&=\delta_{\Lambda\Lambda'} \left(H(x,\xi, t) +\frac{t}{4 M^2}E(x,\xi, t)\right)+\delta_{\Lambda, -\Lambda'}\, \left(\frac{\Delta_x}{2M} \right)\, \left(H(x,\xi, t) +E(x,\xi, t)\right)\quad.
\label{eq:def_gpd}
\end{eqnarray} 
%
%
In particular, when setting $\Delta_y=0$ without loss of generality, the non-flipping terms come from $\chi^{\dagger}\mathbb{I}\chi$ contributions from the nucleon spinor part, while the flipping ones come only from $\chi^{\dagger}\sigma^y\chi$ contributions. So,  in the bag kinematics, Eqs.~(\ref{eq:quark_sym_kin}), with $\Delta_y=0$, the GTMD structure is the following
\begin{eqnarray}
&&\int d^2 \vec{k}_T\left(2F_{13}-F_{11}\right)(x, \xi, \vec{k}_{T}^2, \vec{k}_{T}\cdot\vec{\Delta}_{T}, t)
= H(x,\xi, t) + E(x,\xi, t)\quad,
\label{eq:GM_bag}
\end{eqnarray}
where we have only considered $T$-even distributions. In terms of HAs, it reads,
%
%
\begin{eqnarray}
\left(2 F_{13}-F_{11}\right)^u(x,\xi, k_T^2, \vec{k}_T\cdot \vec{\Delta}_T, t)&=&\frac{2M}{\tilde{\Delta}_1} \left(A_{++,-+}^u-A_{--,+-}^u+A_{+-,--}^u-A_{-+,++}^u\right)/4\quad,\nonumber\\
&=& \frac{4}{3}C_{\mbox{\footnotesize bag}}\frac{\left |\Phi_2\left (x{\bar M}-\frac{\omega}{R}, \, \vec{k}_T \right) \right |^2  }{\left|\Phi_3(\frac{\vec{\tilde{\Delta}}}{2}) \right|^2}\, \frac{2M}{\tilde{\Delta}_x}\,\nonumber\\
&\times& \left[
k_{3x} \,t_0(k') \frac{t_1(k_3)}{k_3} -k'_{x} t_0(k_3) \frac{t_1(k')}{k'}+ \left(k_{3x} k'_z-k'_{x}k_{3z}\right) \,\frac{t_1(k_3)}{k_3}\frac{t_1(k')}{k'} \right ]\quad,\nonumber\\
\label{eqn:HplE}
\end{eqnarray}
%
%
and $-1/4$th of the $u$-distribution for the $d$ quark.
Beside the forward sum rules, the chiral-even twist-$2$ GPDs are related to the electric and magnetic Form Factors through their first Mellin moment. Results for the electric and magnetic form factors in the MIT bag model have been extensively studied, {\it e.g.}~\cite{Ji:1997gm, Benesh:1987ie,Betz:1983dy}. A poor comparison with data is usually achieved for large values of $-t$.
In the MIT bag, there is no free parameter. However, as suggested by the authors of Ref.~\cite{Ji:1997gm}, it is possible to account for boost effects through an effective momentum transfer, $\eta \vec{\Delta}$. Values of $\eta$ as small as $0.35$ are preferred by the data. We have compared our result to the one obtained by Ref.~\cite{Ji:1997gm}, in the case of an unprojected, {\it i.e.} without taking into account the Peierls--Yoccoz correction,  and with $\eta=1-\epsilon_0/M$. Though it is not the favored value for the boost correction, it is the most physical one. Our aim is not to agree with phenomenology, but rather to given insights on the mechanisms that generate the quark OAM in a simple confined system. Our result also quantitatively agrees with Ref.~\cite{Benesh:1987ie}. We notice an improvement of the $-t$ slope for the calculation using the Peierls--Yoccoz correction with respect to the unprojected results.

From the result in Eq.~(\ref{eqn:HplE}), we can confirm that the second Mellin moment leads to a total Angular Momentum~\cite{Ji:1996ek}  of $0.5$, with less than $1\%$ error, from the form factor calculation~\cite{Ji:1997gm}. As explained in the latter reference, it is an expected results since the bag model contains no explicit gluons.
\\

\begin{figure}
\includegraphics[width=8.cm]{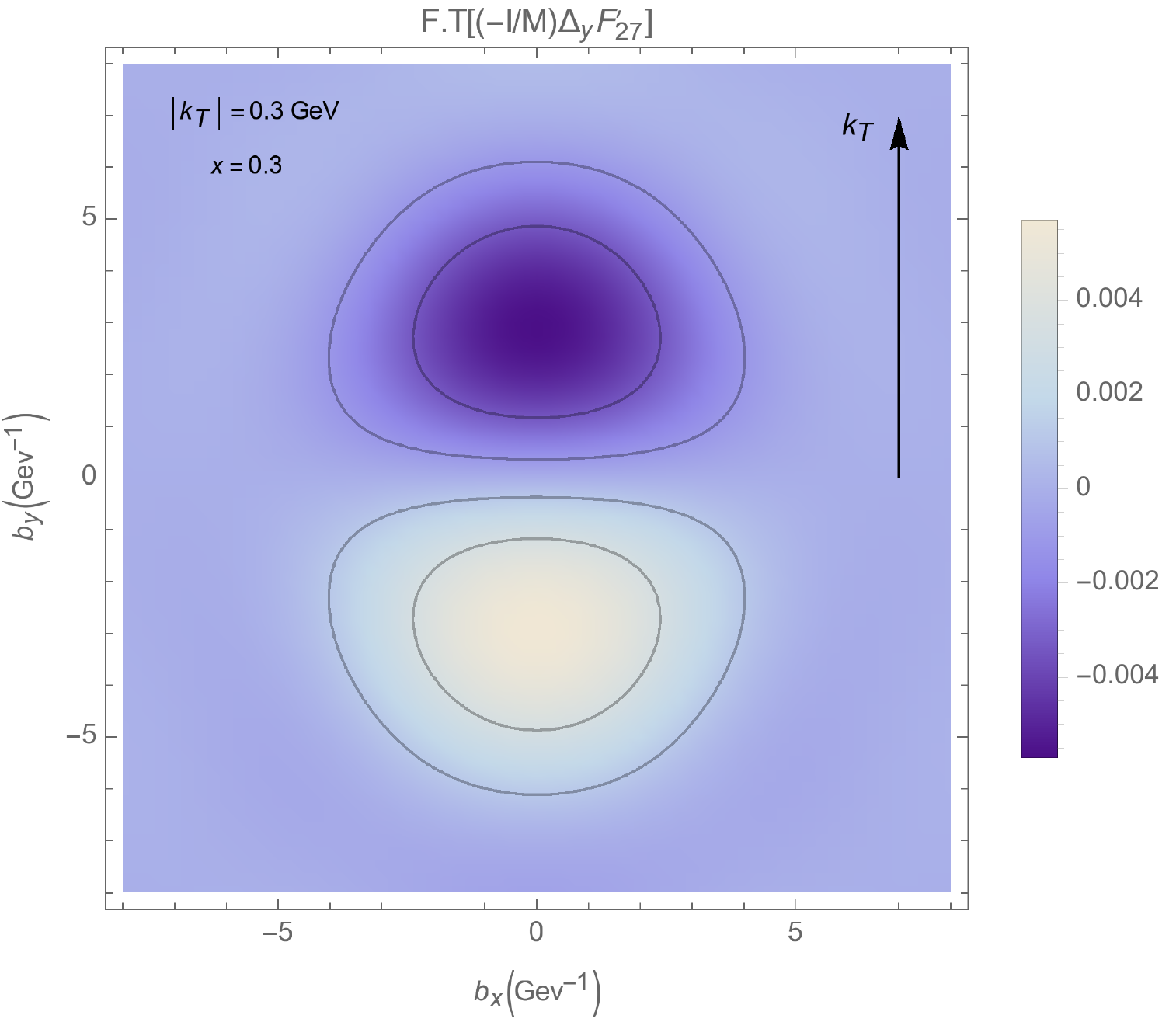}
\includegraphics[width=8.cm]{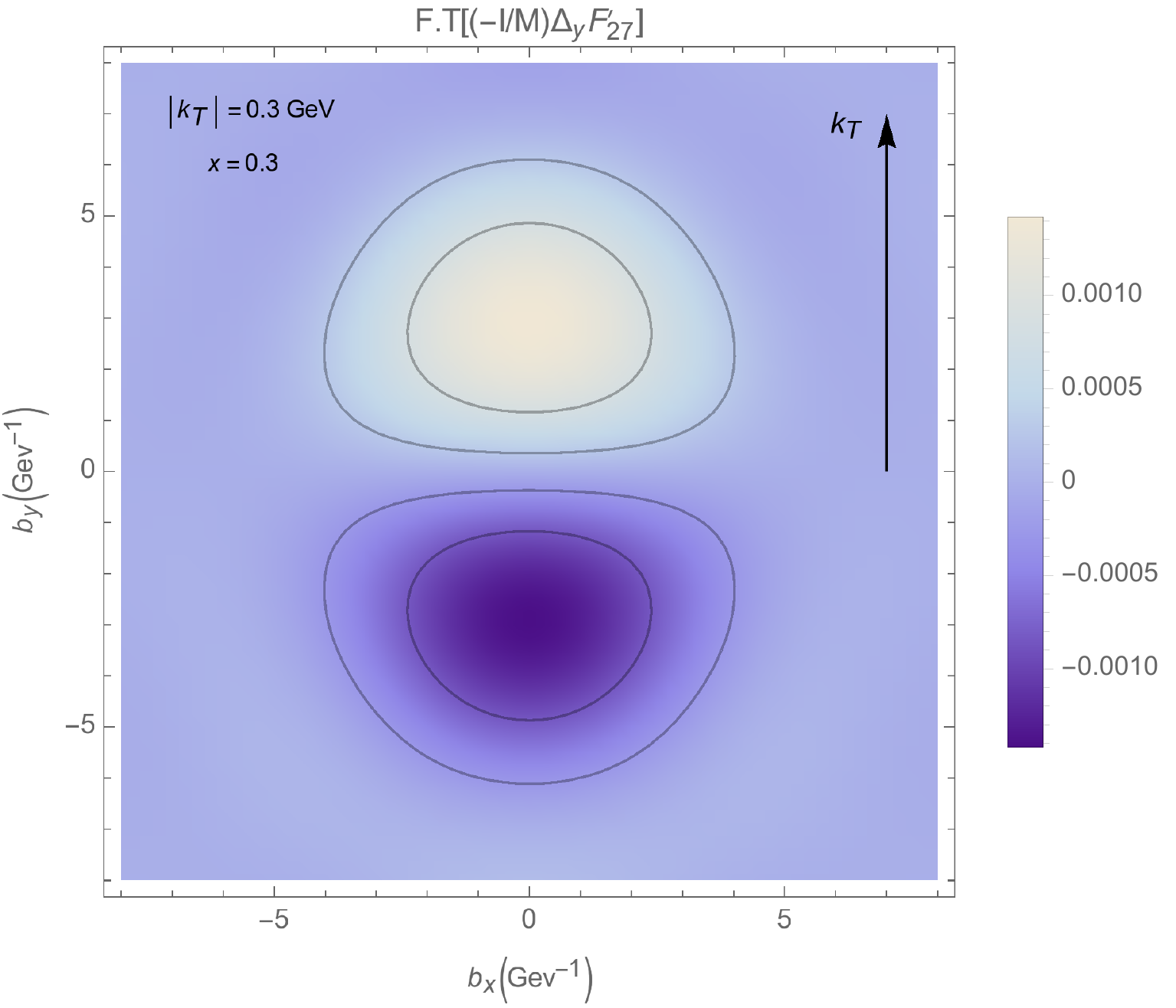}
\caption{Fourier Transform of $F_{27}'$ with the PY projection for $x=0.3$ and $\vec{k}_T=(0.3$ GeV)$ {\hat e}_y$. Up distribution on the {\it l.h.s} and down on the {\it r.h.s.}. Both are proportional as explained in the text.}
\label{fig:f27_FT}
\end{figure}

Next we  calculate the twist-$3$ GPD. The helicity amplitudes for twist-$3$ are defined using chirallity--helicity properties of bad components~\cite{Courtoy:2013oaa,Rajan:2016rlx} %
\begin{eqnarray}
A^3_{\Lambda'\lambda'^{\ast},\Lambda \lambda}
&=&\int\frac{dz^-\, d^2\vec{z}_T}{(2\pi)^3}\, e^{ixP^+z^--i\vec{k}_T\cdot\vec{z}_T}\langle p',\Lambda'|{\cal O}_{\lambda'^{\ast}\lambda}(z)|p,\Lambda\rangle |_{z^+=0}\quad,\nonumber\\
\end{eqnarray}
with
\begin{eqnarray}
{\cal O}_{-^{\ast}+}(z)&=& \bar{\psi} \left(-\frac{z}{2}\right)\; (\gamma^1-i\gamma^2)\, (\mathbb{I}+\gamma_5)\; \psi \left(\frac{z}{2}\right)\quad,\nonumber\\
{\cal O}_{+^{\ast}-}(z)&=& \bar{\psi} \left(-\frac{z}{2}\right)\; (\gamma^1+i\gamma^2)\, (\mathbb{I}-\gamma_5)\; \psi \left(\frac{z}{2}\right)\,;\nonumber\\
\end{eqnarray}
and, equivalently, taking the opposite helicity for $A^3_{\Lambda'\lambda'^{\ast},\Lambda \lambda} \to A^3_{\Lambda\lambda,\Lambda' \lambda'^{\ast}}$.
%
%
%
We will focus only on the forward limit. In this case, the relation between GTMDs and GPDs reads
\begin{eqnarray}
\tilde{E}_{2T}(x, 0, 0)&=&\lim_{t\to0} \, \int d^2\vec{k}_T (-2) 
\left [ \frac{\vec{k}_T\cdot\vec{\Delta}_T}{\Delta_T^2}\, F_{27}(x,\xi, k_T^2, \vec{k}_T\cdot \vec{\Delta}_T, t)+\, F_{28}(x,\xi, k_T^2, \vec{k}_T\cdot \vec{\Delta}_T, t)\right]\,,\nonumber\\
\end{eqnarray}
for which the twist-$3$ HAs are
\begin{widetext}
\begin{eqnarray}
\frac{k_x-ik_y}{P^+}F^u_{27}+\frac{{\Delta}_x-i{\Delta}_y}{P^+}F^u_{28}
&=& -\frac{1}{2}\left( A^3_{+ -^{\ast},++}-   A^3_{+ -,++^{\ast}} - A^3_{- -^{\ast},-+}+  A^3_{--,-+^{\ast}}\right)\quad,\label{eqn:f278}\\
&=&-\frac{8}{3}C_{\mbox{\footnotesize bag}}\frac{\left |\Phi_2\left (x{\bar M}-\frac{\omega}{R}, \, \vec{k}_T \right) \right |^2  }{\left|\Phi_3(\frac{\vec{\tilde{\Delta}}}{2}) \right|^2}\,\left[ 
(k_x-ik_y)\,\left(t_0(k_3)\, \frac{t_1(k')}{k'}- t_0(k')\, \frac{t_1(k_3)}{k_3}\right)\right.\nonumber\\
&&\left.+ \left(\tilde{\Delta}_x-i\tilde{\Delta}_y\right)\,\frac{1}{2}\left(t_0(k_3)\, \frac{t_1(k')}{k'}+ t_0(k')\, \frac{t_1(k_3)}{k_3}\right)\right]\quad.
\end{eqnarray}
The $d$-quark distribution corresponds to $-1/4$th of the $u$-distribution as well, as dictated by $SU(6)$.
Hence, summing the flipping GPD terms to $\tilde{E}_{2T}$, we obtain, in the forward limit,
\begin{eqnarray}
  F_{27}'^u(x, \xi, |k_T|, k_T\cdot \Delta_T, t)&=&-\frac{8}{3} C_{\mbox{\footnotesize bag}}\,\frac{\left |\Phi_2\left (x{\bar M}-\frac{\omega}{R}, \, \vec{k}_T \right) \right |^2  }{\left|\Phi_3(\frac{\vec{\tilde{\Delta}}}{2}) \right|^2}\,
M\, (x M-\epsilon)\frac{t_1(k_3)}{k_3}\,\frac{ t_1(k')}{k'} \, \quad,
   \label{eq:g2}
\end{eqnarray}
\end{widetext}
\begin{eqnarray}
  G_2(x, \xi, t)&=&\int d^2 \vec{k}_T \,F_{27}'(x, \xi, k_T, \vec{k}_T\cdot \vec{\Delta}_T, \vec{\Delta}_T^2)\nonumber\quad.
  \end{eqnarray}
  In Fig.~\ref{fig:f27_FT}, we show the Fourier transform of the GTMD combination that we have labeled $ F_{27}'$,
  \begin{eqnarray}
\mbox{F.T.}\left[- i \epsilon_T^{ij}\cfrac{k\Delta_j}{M} \, F_{27}'(x, 0, \vec{k}_T^2, \vec{k}_T\cdot \vec{\Delta}_T,  \vec{\Delta}_T^2)\right]
&=& -\frac{\epsilon_T^{ij}}{M}\frac{\partial}{\partial b_T^j} {\cal F}_{27}'(x, 0, \vec{k}_T^2, \vec{k}_T\cdot \vec{b}_T,  \vec{b}_T^2)\quad,
\end{eqnarray}
where, obviously, the calculation was done setting $\Delta_y\neq~0$.
Our results agrees with the evaluation of Ref.~\cite{Courtoy:2014bea}, in the reggeized quark-diquark picture~\cite{Goldstein:2010gu}.

\begin{figure}
\includegraphics[width=8.cm]{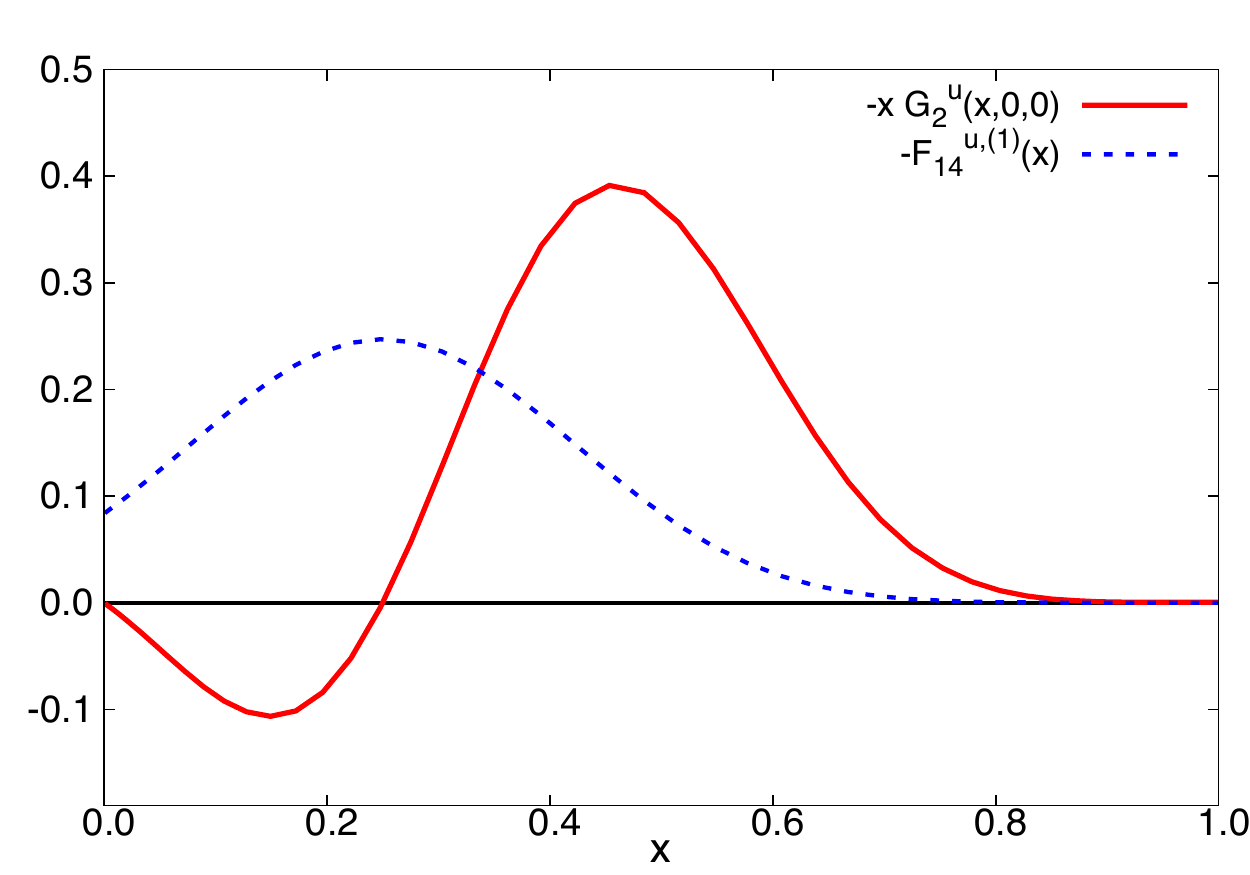}
\caption{The up quark second Mellin moment of $ G_{2}$ in the forward limit, Eq.~(\ref{eq:g2}), compared to $F_{14}$ as in Fig.~\ref{fig:f14}.}
\label{fig:g2comp}
\end{figure}

\begin{figure}
\includegraphics[width=8.cm]{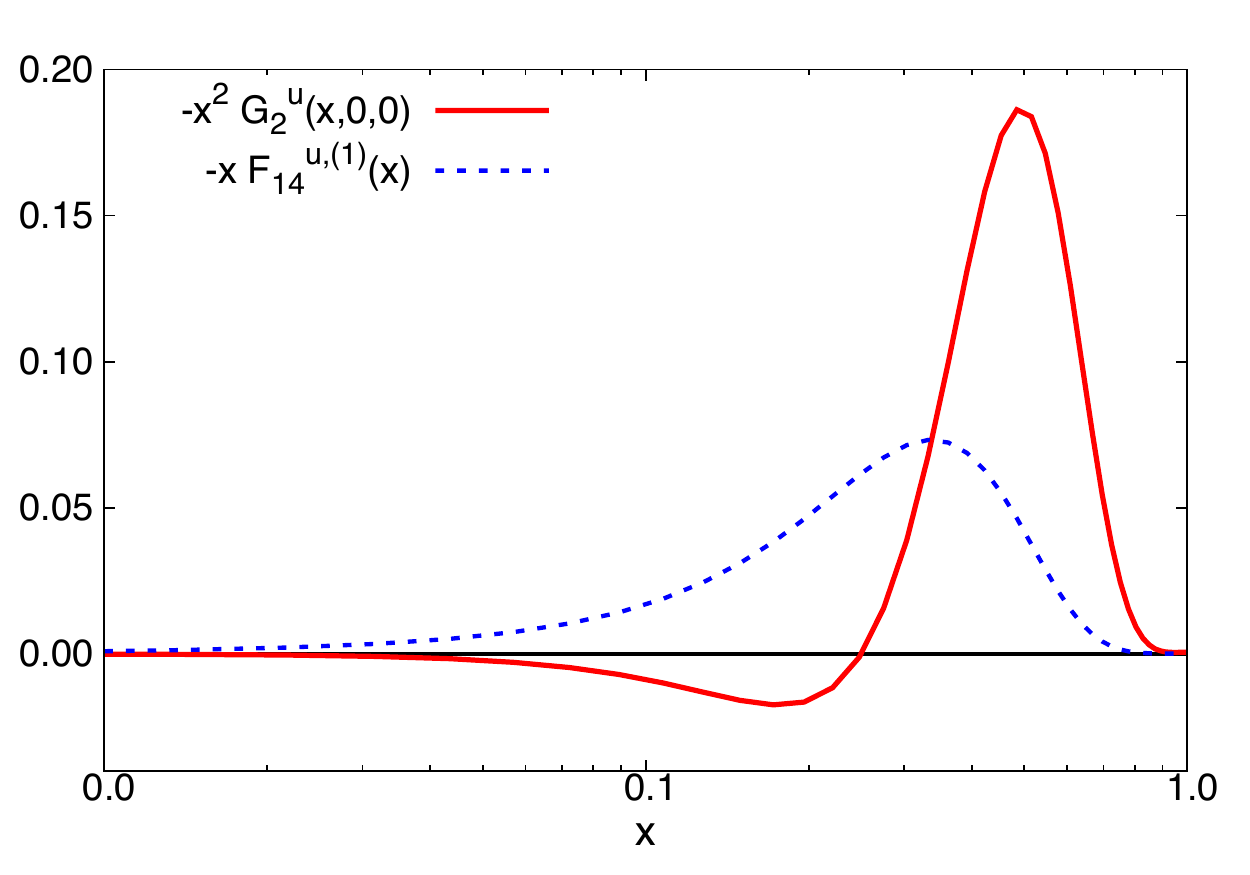}
\caption{The up quark third Mellin moment of $ G_{2}$ in the forward limit, Eq.~(\ref{eq:g2}), compared to $x F_{14}$. 
 }
\label{fig:g2compvicente}
\end{figure}

In Figs.~\ref{fig:g2comp} and \ref{fig:g2compvicente}, we show the behavior of the twist-3 GPD in the forward limit compared to the GTMD $F_{14}$. 
There exists a previous analysis of the OAM by subtraction in the MIT bag model~\cite{Scopetta:1999my}. The result of that reference is based on a previous calculation of GPDs of Ref.~\cite{Ji:1997gm}. While our present analysis is similar, we obtain a slightly qualitatively different behaviour for $G_2$ in the forward limit, {\it i.e.} the solid red curve of Fig.~\ref{fig:g2compvicente}, to be compared with Fig.~2 of that reference.
The discrepancy between our result, explicitly calculated from the twist-3 operator, and the result of Ref.~\cite{Scopetta:1999my} resides in the choice of the approximations. Our calculation is fully unboosted.
\\

%
\section{Sum Rules}

We now have all the ingredients to check the OAM sum rule proposed in Ref.~\cite{Ji:2012sj,Rajan:2016tlg} for integrated densities.

First, using Eq.~(\ref{eq:f14dens}), the quark OAM obtained through the GTMD way with a straight gauge link gives
\begin{eqnarray}
{\cal L}^u_{FS}&=&-\int_{-1}^1 dx\,F_{14}^{u, (1)}(x)=0.139820\quad.
\label{eq:ljm}
\end{eqnarray}
The integral lower limits allows to account partly for spurious support problems but does not qualitatively affects the results, see \cite{Angel:2016}.

On the other hand, the GPD definition of the quark OAM gives
\begin{eqnarray}
 -\int dx\, x  G_2(x, 0, 0)&=&L^q\quad,
 \end{eqnarray}
 which, in the present model calculation, gives
\begin{eqnarray}
 -\int dx\, x  G_2^u(x, 0, 0)
 &=& 0.139817\quad.
 \label{eq:g2_SR}
\end{eqnarray}

We see that the sum rule,
\begin{eqnarray}
-L_q^{\rm Ji} = \int\limits_{-1}^{1} dx F_{14}^{(1)} = \int\limits_{-1}^{1}  dx\, x \, G_2\simeq-0.105\;,
\label{eq:sumruleint}
\end{eqnarray}
proposed in Ref.~\cite{Rajan:2016tlg} is fulfilled within an error of less than $1$\textperthousand, as can be seen comparing Eq.~(\ref{eq:g2_SR}) with Eq.~(\ref{eq:ljm}). 
We notice that this result is confirmed, but at the $\%$ level, without the correction of translation invariance as well. The integration over the Bjorken variable is taken from $-\infty$ to $+\infty$ to make up for the remnant of the support problem. This correction is of less than $1\%$.

Notice that, in the bag model, the sum rule  Eq.~(\ref{eq:sumruleint}) is not only valid for the sum over the flavor but also per flavor, which is in contrast with the result in the light-cone constituent quark or even the chiral quark soliton models~\cite{Lorce:2011kd}.
\\

\begin{figure}[t]
\includegraphics[width=9.cm]{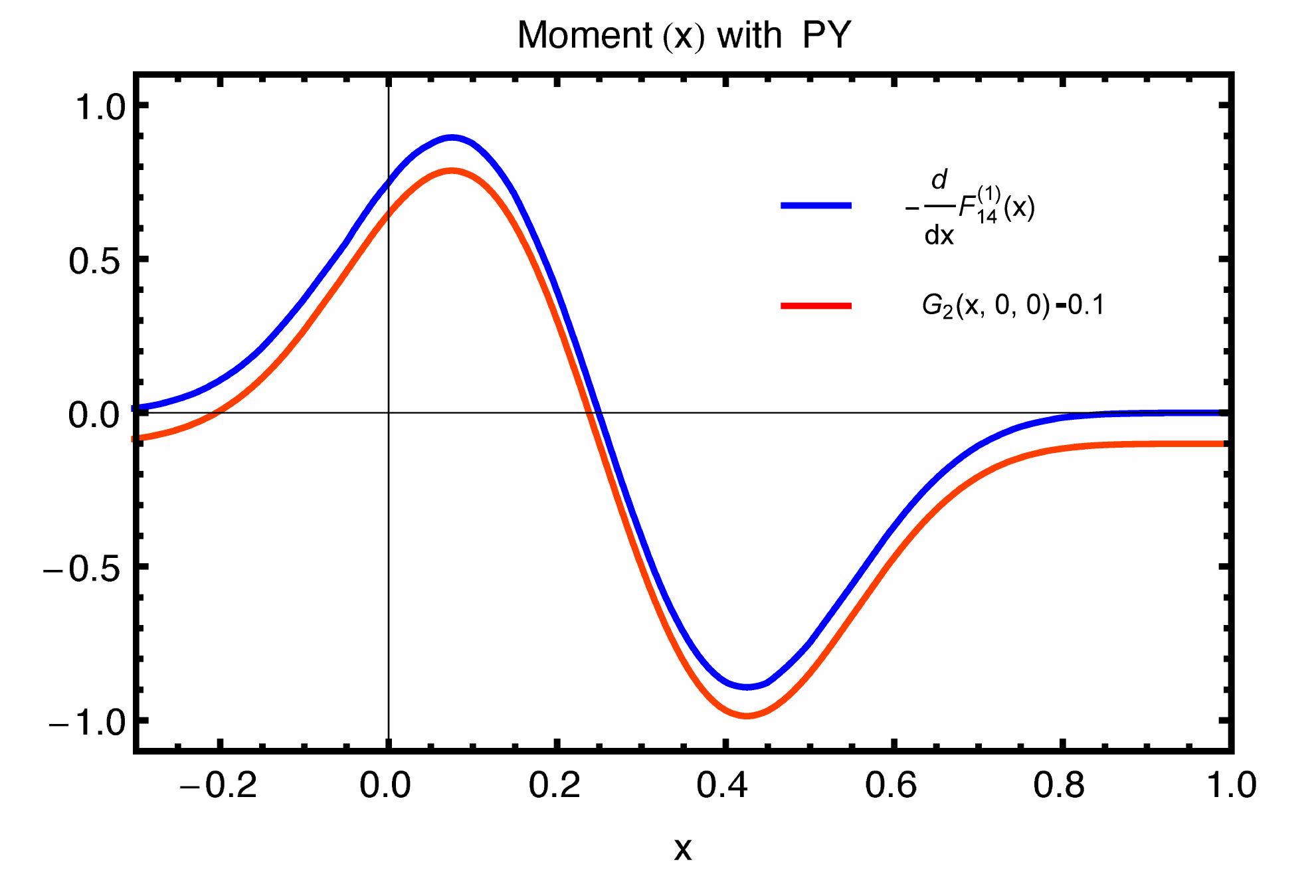}
\caption{The derivative of $ F_{14}$ {\it w.r.t} $x$  compared  $G_2$ in the forward limit, Eq.~(\ref{eq:g2}). Notice that the twist-3 GPD has been shifted to avoid the superposition of the two curves. }
\label{fig:g2WW}
\end{figure}

The main achievement of Ref.~\cite{Rajan:2016tlg} consists in the upgrade of the previous sum rule at a density level. The Lorentz Invariant Relations (LIR) derived in that reference lead to the following expression,
\begin{eqnarray}
\frac{d}{dx} F_{14}^{(1)}(x) &=&   - G_2(x, 0, 0)\qquad.
\end{eqnarray}
From Fig.~\ref{fig:g2comp}, it is straightforwardly deduced that the shape of both densities indeed respect this relation. In Fig.~\ref{fig:g2WW}, we show that it is exactly satisfied within the MIT bag model including PY corrections.

\begin{figure}
\includegraphics[width=8.cm]{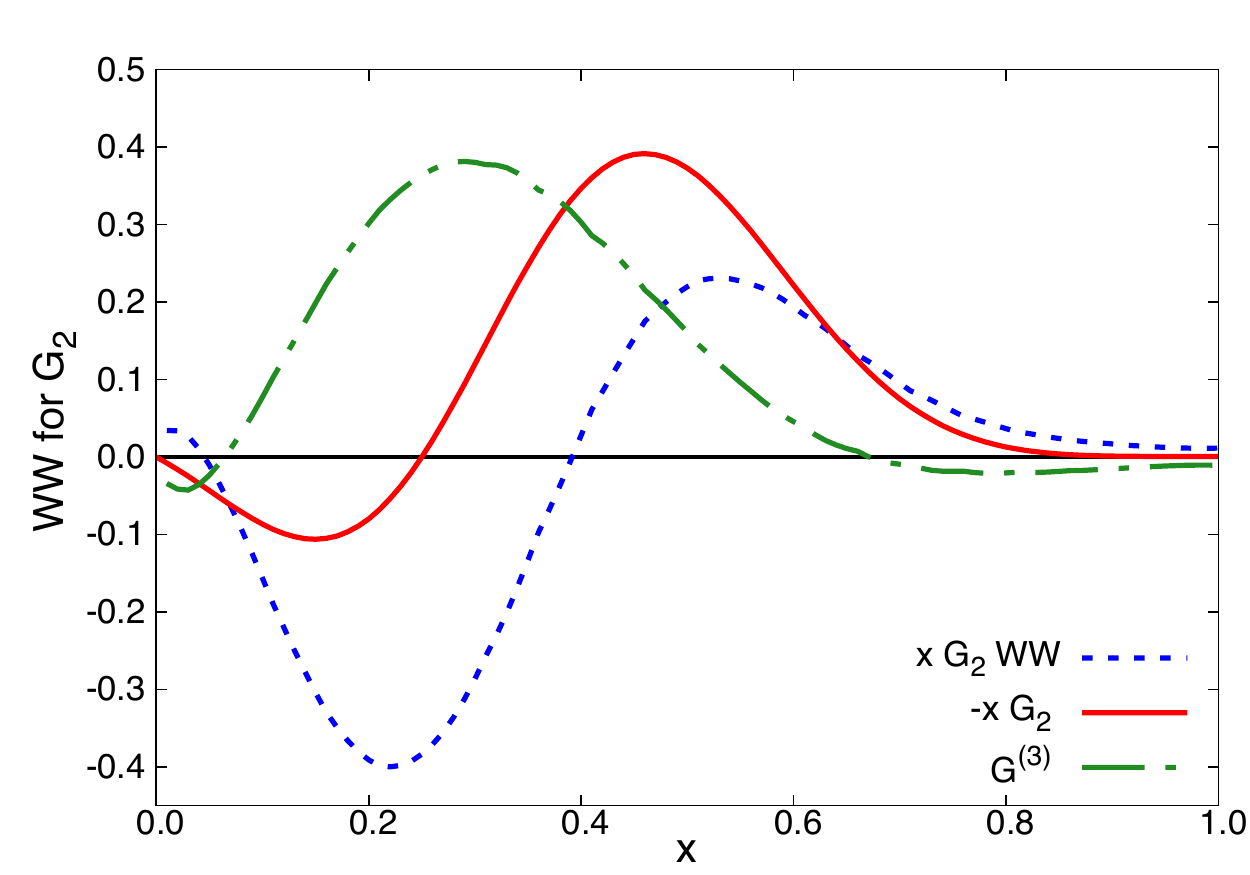}
\caption{The components of Eq.~(\ref{eqn:ww_g2}). The red curve is similar to the red curve of Fig.~\ref{fig:g2comp} though without the PY projection. The genuine part, green curve, is obtained by subtraction.}
\label{fig:wwbla}
\end{figure}

Another way of rewriting the sum rule at the density level follows from the Equations of Motion (EoM) of QCD~\cite{Rajan:2016tlg}
\begin{widetext}
\begin{eqnarray}
-x \, G_2(x, 0, 0)
&=& x \left[ (H+E)
- \int\limits_{x}^1 \frac{dy}{y} (H+E) - \frac{1}{x} \widetilde{H}
+ \int \limits_{x}^1 \frac{dy}{y^2}  \, \widetilde{H} \right] (x, 0, 0)+ G^{(3)}(x, 0, 0)\quad, \nonumber\\
&=& xG_2^{WW} (x, 0, 0)+ G^{(3)}(x, 0, 0)\quad.
\label{eqn:ww_g2}
\end{eqnarray}
\end{widetext}
The WW contribution is directly calculated in the bag model, see Ref.~\cite{Angel:2016} for the expression for $\tilde{H}$. In Fig.~\ref{fig:wwbla}, we show the contribution from the resulting kinematically reducible twist-3 and the genuine quark-gluon interaction counterpart. The latter is calculated by subtraction.
\\


\section{Genuine twist-3 calculation}

Thanks to the characteristics of the MIT bag model, it is possible to calculate directly the genuine twist-3 contribution from the operator without major complications. Since there are no explicit gluon fields in the MIT bag, the only constraint for higher dimensional operators comes from the EoM. We follow the steps of Ref.~\cite{Jaffe:1990qh}.

In the MIT bag, the EoM
\begin{eqnarray}
i\slashed \partial(\xi)(\xi) \, \psi&=& \delta (\xi-R)\psi(\xi)\quad,
\label{eq:bageom}
\end{eqnarray}
 is intimately related to the confinement mimicked through the boundary condition. Refering to the OPE definition of the tower of twist-2 and twist-3 GPDs~\cite{Belitsky:2000vx}, the previous equation constrains the tower of operators to give the following expression of operators in the model
 \begin{widetext}
\begin{eqnarray}
{\cal O}_{3, \mbox{\tiny gen}}^{\sigma\mu_1\cdots \mu_n}&=& \frac{i^n}{2} \Sym_{\mu_1 \dots \mu_n} \left\{\bar\psi(0)\delta (\xi-R) \sigma^{\sigma\mu_1} \, 
\!\stackrel{\leftrightarrow}{\cal \partial}^{\mu_1}
\dots
\!\stackrel{\leftrightarrow}{\cal \partial}^{\mu_n}
\psi (0)+
\bar\psi(0) \sigma^{\sigma\mu_1} \, 
\!\stackrel{\leftrightarrow}{\cal \partial}^{\mu_1}
\dots
\!\stackrel{\leftrightarrow}{\cal \partial}^{\mu_n}
\delta (\xi-R)\psi (0)\right\}
\quad,
\end{eqnarray}
\end{widetext}
The Dirac structure found in the above equation is similar to the one highlighted in Ref.~\cite{Rajan:2016tlg}. Interestingly, it also coincides with the PDF $g_2$~\cite{Jaffe:1990qh}. Our result will then only differ by the Lorentz structure of the twist-3 GPD. The latter was proposed in Ref.~\cite{Rajan:2016tlg},
\begin{equation}
\widetilde{\cal M}
 = 2M \frac{\Delta_{T}^{i} }{\Delta_{T}^{2} }
\int d^2 k_T \, \left[ {\cal M}^i_{++} - {\cal M}^i_{- - }  \right] \,,
\label{gen3}
\end{equation}
where, in the MIT bag with no explicit gluons and using EoM Eq.~(\ref{eq:bageom}),
%
\begin{eqnarray} 
{\cal M}^i_{\Lambda \Lambda'} &=& \frac{1}{2}
\int \frac{d z^- d^2 z_T}{(2 \pi)^3} e^{ixP^+ z^- - i k_T\cdot z_T}
\langle p^{\prime } ,\Lambda^{\prime } \mid \overline{\psi}(-z/2) 
i \sigma^{i+} \gamma_5   \delta (z/2-R) \psi (z/2)  \mid p, \Lambda \rangle \mid_{z^+=0} \,.\nonumber\\
\label{eq:mgen}
\end{eqnarray}

 While it was noted that the $k_T$-integral must vanish in the limit $\Delta_T\to 0$ to keep $\widetilde{\cal M}$ regular, it does not happen in the present model. It is clear from Ref.~\cite{Belitsky:2001ns} that the transverse index of the Dirac structure must be contracted by $\Delta_T^i$. The same result was found in \cite{Rajan:2016tlg} through the EoMs. That dependence suggests that the genuine twist-3 diverges at the limit of zero momentum transfer. We believe it is an artifact of the model combined with the scarcity of information about structure functions related to genuine twist-3 parton distributions. Here, for the present model evaluation, we explore a non-divergent structure for the bag model,
\begin{equation}
\widetilde{\cal M}_{\mbox{\tiny ND}}
 = 2 M \frac{\Delta_{T}^{i} }{P\cdot q}
\int d^2 k_T \, \left[ {\cal M}^i_{++} - {\cal M}^i_{- - }  \right] \,,
\label{gen3me}
\end{equation}
with $P\cdot q=-2 \xi \bar{M}^2$. We present results for small values of both $\xi$ and $\Delta_T$, for which the expression is still convergent.
\\

The calculation of the correlator in Eq.~(\ref{eq:mgen}) at the boundary requires little adjustments of the definitions.
At the boundary, the Fourier transform of the bag wave function is expressed as
\begin{eqnarray}
 t_{i, R}(k) &=& \frac{j_i(kR)j_i(\omega)}{R}\quad,
\end{eqnarray}
where the dimensions have been adapted adequately to keep the overall constant unchanged.
The expression for the genuine twist-3 contribution then reads, in the forward limit $k_3=k'\equiv k$,
\begin{eqnarray}
\left[ {\cal M}^i_{++} - {\cal M}^i_{- - }  \right] 
&=&\frac{C_{\mbox{\footnotesize bag}}}{R}
\left[
t_0(k)\, j_0(\omega)j_0(kR)+\frac{k_z}{k}( t_0 (k)\, j_1(\omega)j_1(kR)+t_1(k) \, j_0(\omega)j_0(kR) )
+\frac{k_z^2}{k^2} t_1(k) \, j_1(\omega)j_1(kR)
\right] \,.\nonumber\\
\label{eq:genbag}
\end{eqnarray}

\begin{figure}
\includegraphics[width=8.cm]{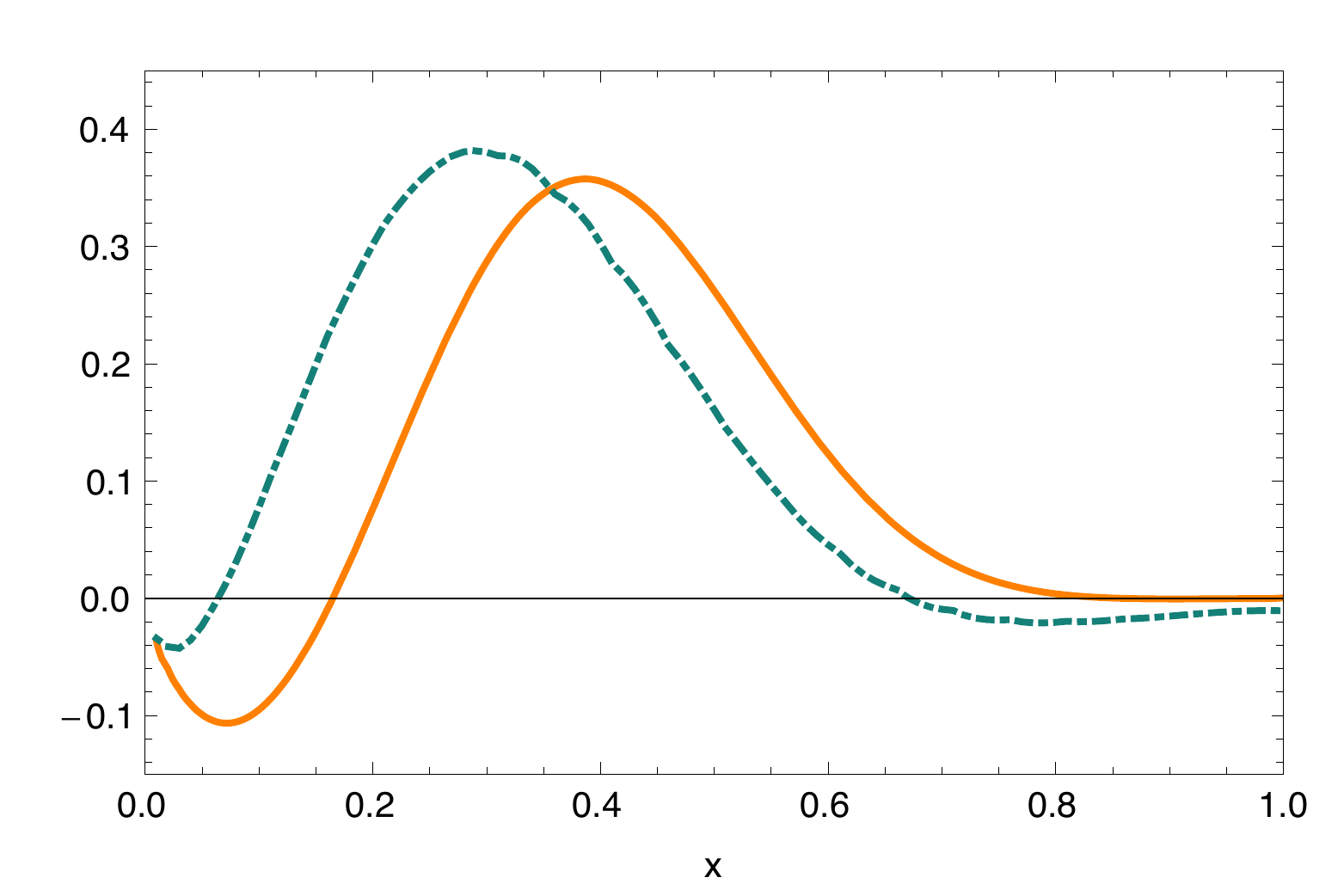}
\caption{Comparison of the genuine twist-3 contribution to $G_2$ obtained by subtraction (in dot-dashed green) and obtained through the interaction-dependent correlator (in orange). }
\label{fig:t3comp}
\end{figure}

Combining Eq.~(\ref{gen3me}) and Eq.~(\ref{eq:genbag}), we are able to calculate the genuine twist-3 contribution to $G_2$ given by
\begin{equation}
G^{(3)} = -\widetilde{\cal M}
+x\int_{x}^{1} \frac{dy}{y^2 } \widetilde{\cal M}\quad.
\end{equation}
The result is depicted in Fig.~\ref{fig:t3comp} including the PY projection, compared to the genuine twist-3 obtained in the previous Section. We notice that the agreement is not perfect, in particular there is a clear shift in $x$ of the result obtained through the interaction-dependent operator. However, the behavior is clearly similar and the quality of the agreement corresponds to that of Ref.~\cite{Jaffe:1990qh}.

\section{Conclusions}

We have presented a model evaluation of the two distribution functions related to the quark Orbital Angular Momentum. On the one hand, we have performed the calculation of the GTMD $F_{14}$ within the MIT bag model corrected via the Peierls--Yoccoz projection. The obtained result agrees with previous evaluations in models for the proton. On the other hand, the subleading GPD $G_2$ has been evaluated in the same model. There are no other model calculation with explicitly given $x$-behavior to compare our result to. 

The main goal of the present paper was to corroborate the sum rules at the density level proposed in Ref.~\cite{Rajan:2016tlg}. The first relates the derivative of the GTMD {\it w.r.t.} $x$ to the GPD $G_2$. We have shown that it is fully satisfied within our model approximations. 
The second extracts the Wandzura--Wilczek kinematical part in terms of known GPDs leaving only the genuine quark-gluon interaction unknown. The benefits of the characteristics of the bag model come in its dynamics: the boundary conditions mimic the role played otherwise by confining gluons. As a first step, we have evaluated the kinematically reducible contributions and deduced the expected behavior of the genuine twist-3. Then, thanks to the equation of motion of the bag, it has been possible to directly calculate the quark-gluon interacting terms. The result in both approaches is slightly different, especially due to a shift in $x$. 

The results obtained using a model that only simulate some of the aspects of QCD are satisfactory. There was no ambition of delivering proper numbers for the sum rule, but mainly to verify them in a model for the proton. While it is believed that the absence of explicit gluons renders automatically the equality between the Orbital Angular Momentum \`a la Ji and \` a la Jaffe-Manohar, it is true with a similar choice of path for the gauge-link in both evaluation. We believe the results would change qualitatively once a staple gauge-link is included in the definition of $F_{14}$.
Further studies should include more model estimates of the presently discussed sum rule as well as an extension towards the inclusion of the gauge-link in a different gauge choice for the Generalized Transverse Momentum Distribution.
\\

\begin{acknowledgments}
The authors thank useful discussions with S. Liuti, C. Lorc\'e and A.~Rajan as well as  advices and enlightening discussions with V. Vento. 
\end{acknowledgments}

\bibliographystyle{apsrevM}
\bibliography{bag}

\end{document}